\documentclass{article}
\usepackage{graphicx}
\usepackage [english]{babel}
\usepackage [autostyle, english = american]{csquotes}
\MakeOuterQuote{"}
\newcommand{\hl}[1]{{#1}}

\begin{document}

%
% The "title" command has an optional parameter, allowing the author to define a "short title" to be used in page headers.
\title{An Analysis of the Consequences\\ of the General Data Protection Regulation\\on Social Network Research\footnote{To appear in: ACM Transactions on Social Computing}}

%
% The "author" command and its associated commands are used to define the authors and their affiliations.
% Of note is the shared affiliation of the first two authors, and the "authornote" and "authornotemark" commands
% used to denote shared contribution to the research.\author{Andreas Kotsios}
\author{Andreas Kotsios\footnote{Faculty of Law, Uppsala University} \and Matteo Magnani\footnote{InfoLab, Dept. of Information Technology, Uppsala University}  \and Luca Rossi\footnote{IT University Copenhagen} \and Irina Shklovski\footnote{IT University Copenhagen} \and Davide Vega\footnote{InfoLab, Dept. of Information Technology, Uppsala University}}

\date{}

\maketitle

%
% The abstract is a short summary of the work to be presented in the article.
\begin{abstract}
This article examines the principles outlined in the General Data Protection Regulation (GDPR) in the context of social network data. We provide both a practical guide to GDPR-compliant social network data processing, covering aspects such as data collection, consent, anonymization and data analysis, and a broader discussion of the problems emerging when the general principles on which the regulation is based are instantiated to this research area.
\end{abstract}


%
% Keywords. The author(s) should pick words that accurately describe the work being
% presented. Separate the keywords with commas.

%
% A "teaser" image appears between the author and affiliation information and the body
% of the document, and typically spans the page.
%%  \includegraphics[width=\textwidth]{sampleteaser}
%%  \caption{Seattle Mariners at Spring Training, 2010.}
%%  \Description{Enjoying the baseball game from the third-base seats. Ichiro Suzuki preparing to bat.}

%
% This command processes the author and affiliation and title information and builds
% the first part of the formatted document.

\section{Introduction}

In the last decade online social network \hl{platforms} have become a major source of data to study human and social behaviour \hl{\cite{gonzalez2008understanding, bond201261}}. The availability of persistent and searchable traces of human communication on a large scale \cite{Boyd2008} provided new, previously inconceivable opportunities for unobtrusive research, but also raised new questions related to the potential misuse of personal information \hl{\cite{Lazer721,O'Neil:2016:WMD:3002861}}. Following events such as the Cambridge Analytica scandal, and related restrictions on research-related data access established by large social media companies, some Internet researchers have highlighted the necessity and complexity of ensuring that "independent, critical research in the public interest can be conducted while protecting ordinary users' privacy"\footnote{https://medium.com/@Snurb/facebook-research-data-18662cf2cacb}. \hl{For European researchers, this context is further complicated by the recent advent of the General Data Protection Regulation\footnote{Regulation (EU) 2016/679 of the European Parliament and of the Council of 27 April 2016 on the protection of natural persons with regard to the processing of personal data and on the free movement of such data.} (GDPR), which came into force on the 25th of May 2018.} 

\hl{The GDPR is a piece of European legislation regulating how natural persons should be protected with regards to the processing of their personal data. The GDPR applies to processing of personal data in very similar ways in all EU Member States, for all sectors (public or private) and all purposes (commercial and non-commercial). This includes research performed by private companies or public universities and other research institutions}. The regulation has been welcomed as a progressive step towards rectifying the glaring power imbalance in current mass digital data collection by entities that develop, maintain and control access to digital infrastructures. \hl{T}he GDPR has two main goals\footnote{Rec 1-7 GDPR and SOU 2017:50, p78}. The first goal is to protect the fundamental rights and freedoms of the data subjects by creating a protective regiment with regards to the processing of personal data\footnote{In the context of GDPR 'personal data' means any information relating to an identified or identifiable natural person; an identifiable natural person is one who can be identified, directly or indirectly, in particular by reference to an identifier such as a name, an identification number, location data, an online identifier or to one or more factors specific to the physical, physiological, genetic, mental, economic, cultural or social identity of that natural person.}. This is because new technologies and organisational models both in the private and public sector have made it easy to gather, use, combine, aggregate or otherwise process a vast amount of personal data without sufficient controls or oversight. The second goal is to create the optimal conditions so that the free flow of personal data -- in parallel to the free movement of goods and services -- can take place within the EU, supporting the creation of the European Single Market. \hl{The GDPR is intended to provide a way of achieving the free flow of data within the EU while ensuring protection of the fundamental rights and freedoms for individuals.}

The regulation replaces the earlier Data Protection Directive\footnote{Directive 95/46/EC} which, as a directive, was adopted and implemented through different national laws by every EU Member State, resulting in at times confusing patchwork of national regulations. More importantly, the earlier directive did not have any specific focus on research. Rather, the main regulatory mechanisms were codes of conduct and ethical guidelines advocating good practices but rarely systematically codifying these. In contrast, the GDPR explicitly recognizes the particularities of data processing in research through a series of formally specified research exemptions, which have important consequences on the feasibility and lawfulness of social network research projects in practice. This includes the ability to limit and even avoid restrictions on secondary processing and the processing of sensitive categories of data\footnote{Art 6(4) and rec 50 GDPR}, to override the subjects' right to object to processing and erasure as long as relevant safeguards are implemented\footnote{Art 89 GDPR}, and to collect some types of data without consent for some types of processing\footnote{Art 6(1)(e) and (f); rec 47 and 157 GDPR}.

\hl{T}he impact of the new rules on the practice of research is unclear and this is especially relevant to researchers studying social network data (both on- and offline), where for example the subjects participating in a study may provide information about non-participants and the collected data is more difficult to effectively anonymize than in other research fields. This article considers the impact and implications of the GDPR and of the research exemptions built into the law on the activities of researchers engaging in social network analysis in general and in the specific case of online social networks. 

The text of the GDPR is complex and not specifically targeted to researchers but its content will impact research practices in significant ways that depend on the specific research field. On the one hand, universities and other research institutions are of course providing general information about the GDPR to their employees, and have a Data Protection Officer who can be contacted for specific matters. On the other hand, GDPR-compliant strategies need to be instantiated to the specific research problem, and many of these are difficult to interpret without domain-specific knowledge. Therefore, researchers should themselves be aware of the implications of the GDPR and have reflected about it. Several papers have already been published to address this issue and to examine the impact of the GDPR on research in general \cite{Chassang2017,Marelli2018,Schaar2016} and on specific research fields \cite{Penasa2018}. However, none have addressed the implications of the GDPR for social network analysis, where data processing differs from other quantitative approaches. For example, Borgatti and Molina \cite{Borgatti2005} \hl{point out that} respondent anonymity is not an option if we want to know who is talking about whom, which is necessary to define edges in the network. In addition, subjects providing information about their social relations may generate data about individuals not included in the study: a participant mentioning that she often performs some activity with someone may reveal a lot about this other person depending on the type of activity. Another issue is the fact that in social networks it is often possible to identify specific roles based on the network structure, with a limited number of individuals in each role, examples being high-degree and high-betweenness\footnote{Degree and betweenness are so-called centrality measures, that can be used to identify important actors in a network.} nodes as well as other special network configurations. Once these few nodes have been identified, it becomes very simple to connect them to specific individuals using some basic knowledge of the studied organization. For this reason network data is often impossible to fully anonymize. In this article we will also identify specific issues related to data protection emerging when social network analysis is applied to contexts such as the analysis of large-scale networks of social relations derived from social media data.

In the next section we present a brief overview of the GDPR, including the terminology used in the rest of the article. This short section is necessary to make this article self-contained, and it can be skipped by the reader who is already familiar with the main actors, concepts and principles introduced by the regulation. The following section is organized along the main steps and problems of a typical social network research process. We start by discussing approaches to data collection, also highlighting the differences between data collected directly from the data subjects or indirectly, such as through social media Application Programming Interfaces (APIs). We also discuss topics such as consent, data anonymization, profiling, and storage of the networks. We conclude the article with more general considerations about the implications of the GDPR for commercial data controllers as well as for the future of network data repositories which represent important teaching and training tools. We also suggest that the GDPR should be a new important element to be considered in the ongoing discussion about the establishment of a code of conduct for social network research\footnote{\hl {The GDPR applies only to the processing of personal data by entities established in the European Union regardless the place of processing, or in general to processing of personal data of data subjects who are in the European Union, as long as the processing is related to the offering of goods and services in the Union or the monitoring takes place in the Union.} Our presentation will often take the perspective of a European public university, and we will extend the discussion to other cases regulated by this law when relevant. However, the principles defined in the GDPR are worthy of consideration even for researchers outside the European Union processing data from non-EU subjects, as these principles highlight general fundamental issues to be considered when processing personal data.}.

\hl{Please note that our objectives are (1) to provide a general understanding of the impact of the GDPR on social network research for scholars with no background in law, under the assumption that their institutions will be able to fill in the details about local regulations but may be unaware of the aspects more specific to social network data, and (2) to highlight some controversial issues. It is not our goal to provide a detailed technical legal analysis of the GDPR; for this we refer to the relevant current debate and scholarly work\footnote{\hl{See for example Feiler L, Forg\'o N and Weigl M, The EU General Data Protection Regulation (GDPR): A Commentary (Globe Law and Business 2018) and Kuner C, Bygrave L and Docksey C (eds), The Eu General Data Protection Regulation (Gdpr): A Commentary (Oxford University Press 2019).}}. Similarly, we do not provide legal analyses of specific cases, as this is out of scope for the article.}

\section{The GDPR ecosystem: overview and terminology}

Before we proceed to \hl{our analysis of} how the GDPR may affect social network research in practice it is important to make clear the fundamental terms and ideas of this piece of law. The regulation is complex in terms of length (88 pages, 173 recitals, 99 articles), breadth of coverage and depth. Here we present the GDPR concepts and principles that we find are the most relevant for social network analysis research. In the next section we will discuss the (often unclear) role that these concepts and principles can play in the various phases of a social network analysis process. \hl{In the article we will refer to articles in both the GDPR and the recitals. Even though recitals are not part of the operative text of the regulation their role is of great importance since their purpose is "to set out concise reasons for the chief provisions of the enacting terms"\footnote{\hl{LEGAL SERVICE, EUROPEAN PARLIAMENT ET AL., JOINT PRACTICAL GUIDE OF THE EUROPEAN PARLIAMENT, THE COUNCIL AND THE COMMISSION FOR PERSONS INVOLVED IN THE DRAFTING OF LEGISLATION
WITHIN THE COMMUNITY INSTITUTIONS (2015), section 10, available at https://publications.europa.eu/en/publication-detail/-/publication/3879747d-7a3c-411b-a3a0-55c14e2ba732.}}. We will not discuss here the relationship between recitals and articles, but it is worth keeping in mind that when the legal text of the regulation is somewhat ambiguous it will normally be interpreted in light of the relevant recital\footnote {\hl{Klimas, Tadas and Vaiciukaite, Jurate, The Law of Recitals in European Community Legislation, ILSA Journal of International \& Comparative Law, Vol. 15, 2008.}}}.
 
 \begin{figure}[h]
  \centering
  \includegraphics[width=.7\linewidth]{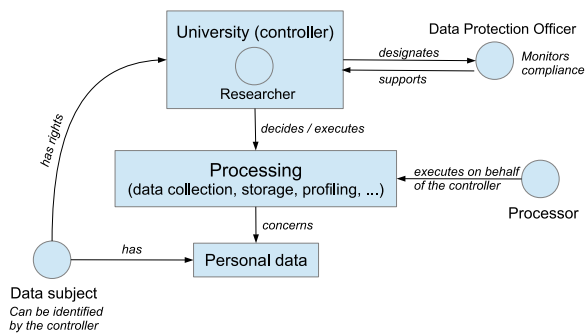}
  \caption{\hl{A typical configuration of the GDPR ecosystem.}}
  \label{fig:gdpr}
\end{figure}

Figure\ref{fig:gdpr} exemplifies the main concepts described in this section in the context of a typical academic research project\footnote{\hl{In Section~\ref{sec:private/public} we discuss the difference between universities and commercial research.}}. The data subject is defined as every individual (natural person) who is identified or may be identified by the controller or third parties, directly or indirectly, by the act of processing her personal data. The natural or legal persons who decide how and why the personal data will be processed are the data controllers while the ones that process the data on behalf of the controller are the data processors\footnote{Art 4(8) and 4(9) GDPR}. 

What constitutes personal data is defined quite broadly as any information that does or may lead to the identification of a natural person\footnote{Art 4(1) GDPR}. The term processing is defined similarly broadly as "any operation or set of operations on personal data or sets of personal data"\footnote{Art 4(2) GDPR}, including data collection. A special type of processing is profiling. With an equally broad definition profiling is defined as "any form of automated processing of personal data evaluating the personal aspects relating to a natural person, in particular to analyse or predict aspects concerning the data subject's performance at work, economic situation, health, personal preferences or interests, reliability or behaviour, location or movements"\footnote{https://gdpr-info.eu/recitals/no-71/}. 

A data protection officer is the person designated by the controller and/or the processor in case the processing a) is carried out by a public authority,  b) contains systematic monitoring of data subjects, or c) consists of large scale processing of special categories of data\footnote{Art 37 GDPR}.  

These definitions, when applied to research, make clear that any research-based processing of social network data that not only directly identify but also possibly may identify (by the same researchers or third parties) individuals will be regulated under the GDPR. 

Typically, in cases where research is conducted under the auspices of a university, the university is considered to be the data controller. While universities are supposed to have organisational measures with regards to the GDPR, the researchers, as employees of the university, who one way or another process personal data as part of their role, are also expected to have an understanding of the GDPR as they design data collection and analysis protocols.

It may happen that other entities assist with the processing, but do not decide the purposes and the manner of it. For example, a researcher may pay individuals who are not employees of the university to perform a data collection. Where these entities are only following the guidelines of the controller, then these external parties can be seen as data processors, where the "processing by a processor shall be governed by a contract or other legal act"\footnote{Art 28(3) GDPR}.  If the research is conducted by a private actor, such as a company, then it is the company that is the controller, and possible external sub-contractors (including researchers) may constitute the processors. It can also be so that a private party and a university can both jointly be regarded as controllers, depending on what agreement exists between these two parties. 

\hl{As we can see here, in case of more than one entity -- meaning different legal entities -- being involved in the processing of personal data, determining the controller(s) is not always an easy task. It is not sufficient that one entity processes data on behalf of another entity, since it is possible that it also processes these data for its own purposes\footnote{Art. 29 Data Protection Working Party (2010) Opinion 1/2010 on the concepts of 'controller' and
'processor', WP 169, p. 9}. It is not sufficient either that a contract may explicitly state the roles of the entities, since they in reality may act in a different way\footnote{Ibid}. The capacity of the controller is based on factual elements and circumstances, on whether or not an entity can -- and does -- indeed determine the \textit{purposes} and the \textit{means} of processing - the "whys" and the "hows". Some questions that help in determining the role of controller are: why is this processing taking place, who initiated it, would an entity process the data if not asked by another entity and if so under what conditions\footnote{Art. 29 Data Protection Working Party (2010) Opinion 1/2010 on the concepts of 'controller' and
'processor', WP 169, p.8}.

It is important to keep in mind that the \textit{purpose} of the processing can be defined only by the controller, meaning that if there are more than one entity defining the purposes of the processing then these entities are joint controllers. The \textit{means}, on the other hand, namely the decision on organizational and technical matters, can be delegated by the controller to a processor. However, "substantial questions which are essential to the core of lawfulness of processing are reserved to the controller"\footnote{Art. 29 Data Protection Working Party (2010) Opinion 1/2010 on the concepts of 'controller' and
'processor', WP 169, p. 15}. If an entity has the power to decide, for example, issues related to the period of storage or access privileges, this entity is, then, de facto a controller concerning this part of the use of data.
}

These complicated distinctions are important to consider and discuss with relevant internal data protection officers because their particular specifications can have an impact on the obligations of the researcher, or alternatively said, on how to comply with the GDPR. As a simple example, it is the controller that is responsible for providing specific information to the data subjects.

Given the diversity of research approaches it is important that researchers understand the particular aspects of the regulation that apply to them. This also means that any collaborative research project must consider what institutional agreements must be made with respect to data processing: a process that may take additional time and must be planned for. 

\begin{table}[]
    \centering
    \begin{tabular}{l l}
    \hline
P1: & lawfulness, fairness and transparency \\
P2: & purpose limitation \\
P3: & data minimization \\
P4: & accuracy \\
P5: & storage limitation \\
P6: & integrity and confidentiality \\
P7: & accountability \\
    \hline
    \end{tabular}
    \caption{The seven basic principles in the GDPR}
    \label{tab:principles}
\end{table}

The regulation introduces seven \hl{important} principles, listed in Table~\ref{tab:principles}, to be followed when processing personal data\footnote{Art 5 GDPR}. (P1) The data must be processed in a lawful, fair and transparent way. (P2) Personal data may only be collected for specified, explicit and legitimate purposes and not further processed in a manner that is incompatible with those purposes. (P3) The data may be processed only if they are adequate, relevant and limited to what is necessary with regards to the purpose of processing. (P4) Only data that are accurate and up to date, to the level that it is possible, may be processed. (P5) Personal data may only be processed for a period that is necessary for the processing and therefore the controllers must create criteria to determine what retention periods are suitable for their purposes. (P6) The controllers must apply technical and organisational measures in order to protect personal data they control against unauthorised and unlawful processing as well as accidental loss, destruction or damage. (P7) The data controllers have the responsibility to be compliant and to be able to demonstrate compliance when needed, which implies that written records must be kept on whether and how the controller is compliant\footnote{\hl{With regards to the principle of accountability we would also like to draw attention to the provisions of art 24 et seq GDPR defining the liabilities, responsibilities and general obligations of the controllers and the processors}}. \hl{These principles have implications for social network analysis research, which are detailed below.}
%\hl{ANDREAS REVIEWER 3 SAYS: \emph{On accountability: this part needs to be revised. The different roles (i.e. obligations and rights) of data controllers, data processors, DPOs, and data subjects are not clear. Who is responsible for what? Moreover, what is the proper role of the researcher under the GDPR?} CAN YOU ADD ONE SENTENCE / MAX TWO ABOUT THIS?: Answer: I added a footnote. The thing is that when we talk about the accountability as a principle it refers only to the controller. So even though I understand what the reviewer wants us to do (to talk about responsibilities and not accountability as a principle) this is not something that we can discuss in this part. We could have another section but I think it will be too much. Moreover the role of the researcher depends on the actions, meaning that we have to see what arrangement exists and what is the factual power of the researcher. I think the only reference should be to which provisions talk about the different responsibilites. COMMENT (Irina - I think one of the problems here is that these principles are just mentioned and we were not explaining how these are connected to the data controller/processor distinctions, etc. We just sort of say that the regulation introduces the seven principles but how these are connected to the stuff above is not at all clear.}

\section{The GDPR in the social network analysis process}

\hl{The principles mentioned in the previous paragraph and in Table~\ref{tab:principles}} need to be instantiated to the specific cases. 
In this section we \hl{will discuss what implications data processing in the context of research has on the practical enactment of the principles. We will also} detail the meaning of these principles when they regulate the processing of social network data, emphasizing the cases where ambiguities arise.

\hl{A summary of the main GDPR-related aspects that should be considered during a social network analysis process, including a list of exemptions that can be applied in research, is presented in Tables~\ref{tab:rules1} and \ref{tab:rules2}. Detailed explanations are presented in the text.}

\begin{table}[]
    \centering
    \hl{
    \begin{tabular}{l p{5cm} l p{7cm}}
    \hline
& General rule & Exemption & Details \\
\hline
1 &
Identify the roles w.r.t.~the GDPR ecosystem (data subjects, controllers, processors,  DPO\dots) and the data flows. &
no & This can be challenging in some cases, consult the DPO if uncertain.
\\
2 &
Identify the nature of the data (personal / non personal / sensitive). &
no &
In case of sensitive data, we can process it (1) if we have explicit consent, (2) if the data was manifestly made public by the data subject (use this carefully), or (3) in case of research purposes, if there are suitable safeguards (e.g., pseudonymization, approval from ethics committee).\\
3  &
Identify explicit and legitimate purposes 
for the processing.  &
yes &
The specification in case of research can be a bit more general (such as the general research area or part of the project, not specific analytical tasks). Some specification of the intended purpose is however necessary.\\
4 &
Identify the lawful basis for data processing. &
no &
Based on national legislation, that is still being produced,  some actors conducting research (e.g. universities) might be assumed to operate in the public interest and therefore the public task basis may primarily be used. Otherwise the consent and legitimate interests bases should be examined.\\
5 &
Define clear temporal limits for data processing. Non-anonymized data can be kept for no longer than is necessary for the purposes of the processing. &
yes &
More extended periods may apply in case of research as long as appropriate safeguards are implemented.\\
6 &
Put in place technical and organizational measures to protect the data, e.g., ensure privacy by design and by default, pseudonymize the data as soon as possible. &
no &
The measures should be proportionate to the aim pursued.\\
7 &
In case of profiling perform a DPIA. &
no &
Consider with the DPO whether a DPIA is necessary.\\
    \hline
    \end{tabular}}
    \caption{\hl{A summary of general rules and exemptions to be considered during the social network analysis process. Column Exemption indicates whether explicit exemptions exist for research, and exemptions (if any) and other considerations are indicated under Details. Abbreviations used in the table: Data Protection Officer (DPO), Data Protection Impact Assessment (DPIA). (Part 1)}}
    \label{tab:rules1}
\end{table}

\begin{table}[]
    \centering
    \hl{\begin{tabular}{l p{5cm} l p{7cm}}
    \hline
& General rule & Exemption & Details \\
\hline
8 &
Inform the data subjects about the collection, purposes and their rights at the time the data is obtained (if obtained directly from the data subject) or within a reasonable period after the data is obtained and no later than a month (if the data is obtained indirectly). &
yes &
For secondary data, providing information is not necessary if the provision of such information proves impossible or would involve a
disproportionate effort,
if this is likely to render impossible or seriously impair the achievement of the objectives of the
processing.\\
9 &
Collect only adequate, relevant and limited data to what is necessary to achieve the purposes of the processing. &
yes &
As the purpose may be specified in less precise terms (see the exception to Rule 3), this rule is also affected. Consider deleting unwanted data as soon as possible, acknowledging and documenting the process.\\
10 &
Data subjects have the right to check if there is data concerning them, and the right to obtain these data. &
no &
Even if not part of the GDPR, national laws may still restrict this right, e.g., secrecy acts.\\
11 &
Data subjects have the right to have the data concerning them erased. &
yes &
Not necessary if it is likely to render impossible or seriously impair the
achievements of the objectives of the processing. National laws may also restrict this right.\\
12 &
Keep data accurate and up to date. &
no &
\\
13  &
If a new purpose emerges, new legal bases for data processing should be identified. &
yes &
If the new purpose is research, further processing is considered to be compatible to the initial purpose.\\
14 &
If the controller changes the purpose of the processing, information must be provided to the data subject prior to this processing. &
yes &
See the exception to Rule 3 about the increased flexibility in the specification of the purpose in case of research.\\
15 &
Keep written records to demonstrate compliance. &
no &
\\
    \hline
    \end{tabular}}
    \caption{\hl{A summary of general rules and exemptions to be considered during the social network analysis process. Column Exemption indicates whether explicit exemptions exist for research, and exemptions (if any) and other considerations are indicated under Details. (Part 2)}}
    \label{tab:rules2}
\end{table}

\subsection{Lawful bases for data processing}

\hl{The first basic principle of GDPR states that the data must be processed in a lawful, fair and transparent way. This means that in order for data to be processed there has to be some lawful basis for doing so.} The GDPR lists six lawful bases for processing of personal data\footnote{Art 6 GDPR}: a) the data subject has given her consent, b) it is necessary for the performance of a contract, c) it is necessary in order for the controller to comply with a legal obligation, d) it is necessary in order to protect individuals' (the data subject's and/or other natural persons') vital interests, e) it is necessary for the performance of a task carried out in the public interest and f) it is necessary for the purposes of legitimate interests pursued by the controller as long as these interests are not overridden by interests and fundamental rights and freedoms of the data subjects. Even though there are no specific lawful bases that are a priori dedicated to research, the three most relevant tend to be (a) the consent of the data subject, (e) the task carried out in the public interest and (f) the legitimate interests of the controller.

\label{sec:consent}

With a long history starting in medical science the practice of informed consent has been for long time the central pillar of research practices involving human subjects \cite{Capron2018}. A key element of the GDPR is that, addressing a growing lack of satisfaction towards the efficacy of informed consent practices \cite{Mantelero2014}, it provides well-defined research exemptions. 

\hl{By examining the GDPR closer, we can notice that when it comes to the question which lawful basis should be used when processing personal data in general, the most important parameters to take into consideration are the identity of the controller \footnote{\hl{Even though the GDPR applies both to public and private actors, the identity of the controller may lead to different outcomes, as we will illustrate later in this paper.}}, the purposes of processing as well as the context of processing. Depending on these parameters, the controller must decide which lawful basis to use for processing. In the case of research, the following lawful bases seem to be the most relevant: the data subject has given her consent for the processing of her personal data\footnote{\hl{Art 6.1(a) GDPR}}, the processing is necessary for the performance of a public task\footnote{\hl{Art 6.1(e) GDPR}}, and/or it is necessary for the purposes of the legitimate interests pursued by the controller\footnote{\hl{Art 6.1(f) GDPR}}. 

In the case where a controller is a university, it may be most suitable to use as a lawful basis that the processing is necessary "for the performance of a task carried out in the public interest"\footnote{\hl{Art 6.1(e) GDPR. See also SOU 2017:50}}. The definition of such tasks is left to Union or Member States law\footnote{\hl{Art 6.3 GDPR}}. There is, however, no need for an explicit statutory provision as long as there is a clear basis in law\footnote{\hl{Rec 41 GDPR}}. Even in cases where no national legislation is introduced with regards to it, it should be accepted that pubic actors, such as universities, may use this lawful basis for processing of personal data\footnote{\hl{SOU 2017:50, p.18}}. Since in many countries universities -- often even private ones -- are considered to be public authorities by law and they act on carrying out tasks of public interest, such as conducting research\footnote{\hl{See for example in the UK the Freedom of information Act 2000 and in Sweden the Higher Education Act 1992:1434}}, the public task basis for processing personal data seems to be the appropriate lawful basis for a social network research project, as long as the processing is necessary for that project\footnote{\hl{According to art 6.2 and 6.3 GDPR as well as rec.~45 GDPR it is stated that Union or Member State law shall define whether the controller performing a task of public interest can be a legal person governed by public law or by private law.}}. This lawful basis puts the onus of ensuring that the rights of the data subject are balanced against the public interest goals of institutions, whose aims presumably are oriented towards the greater good. This basis is not available at all to commercial organizations and research labs -- at least as long as no law provides for that -- who must rely on consent or the legitimate interest basis to process personal data.

With regards to the use of consent\footnote{\hl{It is not the goal of this paper to make an analysis on consent as a lawful basis in general -- for a better understanding we refer to the Article 29 Working Party Guidelines on consent under Regulation 2016/679 -- but it is worth reminding here that a consent for processing of personal data by a data subject has to be freely given, specific, informed and unambiguous.}} as a lawful basis for the processing of data in research, there are some things that have to be taken into consideration. The first one is that even tough this lawful basis can also be used for the processing of personal data by a research project, an entity may use this lawful basis only "if a data subject is offered control and is offered a genuine choice with regard to accepting or declining the terms offered or declining them without detriment."\footnote{\hl{Article 29 Working Party Guidelines on consent under Regulation 2016/679, p. 3}} If this is not possible, something that in social network research -- and research in general -- can be the case, then this lawful basis should not be used \footnote{\hl{Here it is also important to consider that, as we will argue later in this paper, it can be difficult to provide information to the data subjects of a network research project and therefore it can similarly be challenging to provide the possibility for an informed consent.}}. 
Additionally, for public universities since they are public authorities, researchers must always assess whether or not the consent provided by the data subjects is valid, namely if it is indeed freely given or it is given as a product of imbalance in powers between the university and the data subjects\footnote{\hl{In most research projects, this should not be a great issue since data subjects in a network research project do not normally have a direct connection to a university, but it is still worth considering possible problems that may arise.}}. Lastly, one should make a distinction regarding the term consent as developed in the GDPR and as an "ethical standard and procedural obligation"\footnote{\hl{Art 29 Working Party Guidelines on consent under Regulation 2016/679, p 28}}. That means that it can be so that the lawful basis for processing is the public task basis, art 6.1(e) GDPR, but consent is used as an additional safeguard. In this case it is not two lawful bases used for the processing of personal data but only one, the public task base; consent is only a procedural obligation and not the lawful basis provided for in art 6.1(a).

The third possible lawful basis for research is that the processing is necessary for the legitimate interests\footnote{\hl{The meaning of legitimate interests is to be interpreted widely and contain both trivial and more important interests, commercial or societal etc.}} pursued by the controller or a third party. In general this basis is the most flexible one; at the same time a controller should be very careful when using it as a lawful basis. More specifically, the controller should prove that there is some legitimate interest, that there is a necessity to process personal data for this legitimate interest, that the interests and rights of the data subjects are not violated, namely that there is a minimal privacy impact, as well as that the data subject would not be surprised by such a processing or is not likely to object. An important thing to remember here is that it cannot be used as a basis in cases where public authorities are processing personal data in the performance of their tasks\footnote{\hl{Art 6.1 para 2 GDPR}}. Therefore a public university processing data in the performance of their tasks, which include also research activities, should probably avoid basing the processing conducted for a research project on the legitimate interest basis. 

One last thing that we would like to add here is that if the personal data processed are of sensitive character, an entity conducting research - at least an entity,such as a university, that bases their research activities on some piece of legislation - may primarily base the lawful processing of such data on the fact that the processing is necessary for scientific research purposes as long as appropriate measures are deployed according to art 89.1 and the research is based on a law "which shall be proportionate to the aim pursued, respect the essence of the right to data protection and provide for suitable and specific measures to safeguard the fundamental rights and the interests of the data subject" according to art 9.2(j) GDPR\footnote{\hl{Art 9.2(g), namely that the processing is necessary for reasons of "substantial public interest" could also be the basis for lawful processing of sensitive personal data but since art 9.2(j) specifically refers to scientific research purposes, processing that takes place for scientific purposes should be based on the legal ground of art 9.2(j)}}. Following the same argumentation as above we could, however, claim that if the processing is not necessary or if there is still no specific legislation with regards to processing for research purposes, consent could also be used as a lawful ground for such processing, according to art 9.2(a) GDPR\footnote{\hl{Worth mentioning here that in many countries such processing by a university, even if consent is given by the data subject, could take place only after an ethics committee permits it. See also SOU 2017:50 s. 160.}}.}

\subsection{Data collection}

Social networks can be obtained through a wide range of data collection strategies. Below we detail different approaches to data collection for social network analysis and consider the corresponding consequences of the GDPR. It is worth noting that we focus on networks where nodes represent natural persons: the GDPR does not apply when nodes represent companies, or animals, or even deceased persons (even though in this last case Member States may provide for specific rules\footnote{Rec 27 GDPR}).

\subsubsection{Primary vs. secondary data collection and the principle of transparency}

An important conceptual and legal distinction resides in the selection of methods for data collection. For example, there is a significant difference between data that is collected directly from the data subject (e.g. small-medium scale data obtained through surveys) and data that is collected through a third actor (e.g. online social networks obtained from APIs) \hl{without the direct involvement of the data subject}. The difference here is not only in the scale or the nature of the data but in the relation between the data subject and the data controller: two different articles are concerned with providing information to the data subject when the data are collected directly from them\footnote{Art 13 GDPR} and when data about them have not been obtained from them\footnote{Art 14 GDPR}.

In essence, these articles detail some of the ways that the principle of transparency must be put into action. Transparency addresses the right of the data subject to know and understand how the data are being used; it "requires that any information addressed to the public or to the data subject be concise, easily accessible and easy to understand, and [in] clear and plain language [in particular] in situations where the proliferation of actors and the technological complexity of practice make it difficult for the data subject to know and understand whether, by whom and for what purpose personal data relating to him or her are being collected [...]." If personal data are collected, the data subjects should be informed about the collection and its purposes in order to enable them to exercise their rights. Note that this is different from consent (explained in Section~\ref{sec:consent}) but instead refers to the information that must be made available about data processing activities. \hl{Essentially, data subjects should be able to easily find out who might be using their data and for what purposes.}

While making the data subjects aware of the processing and of their rights may seem straightforward when data are collected directly from them, this can become very difficult to accomplish when large networks are obtained from APIs. The potential difficulties to provide information under specific circumstances are acknowledged in the GDPR, where exceptions \hl{for research in particular} are introduced. Article 14 states that  providing information is not necessary if 1) "the data subject already has the information"; or 2) "the provision of such information proves impossible or would involve a disproportionate effort, in particular for [...] scientific or historical research purposes", subject to some safeguards\footnote{Art 89 GDPR}, if providing information "is likely to render impossible or seriously impair the achievement of the objectives of that processing". Article 14 then continues stating that  "[i]n such cases the controller shall take appropriate measures to protect the data subject's rights and freedoms and legitimate interests, including making the information publicly available".

These are some examples of the kinds of research exemptions that are embedded in the GDPR, codifying and specifying research conduct. Both those exemptions apply to social network research based on online data collected from social media platforms assuming that social media platforms have already informed their users through appropriate Terms of Services that their data will be shared with third parties (eg. through APIs) or assuming that the large scale of collected data will require a disproportionate effort to inform all affected data subjects. This is an example of balancing research needs against the derogation of the rights of the data subject. Technically termed "proportionality of the effort", this is a relatively vague concept. The controller, in order to determine whether it is going to be disproportionately difficult to provide the information, must take into consideration the number of data subjects, the age of the data and if there are any appropriate safeguards already adopted\footnote{Rec 62}. If, after this assessment, the controller finds that the effort will be disproportionate, then she has to assess once again whether the effort involved to provide the information to the data subject exceeds the impact and effects on the data subject in the case where the information is not provided. This assessment has to be documented and depending on the outcome the controller may have to take extra measures (such as pseudonymisation or anonymization if possible and appropriate).

As an example, this means that although the research exceptions may not technically require that every single Twitter user of the millions involved in any large-scale Twitter network research be notified that their data are used for research, the logic involved in deciding to collect data and to skip the notification must be formally documented. This documentation must also demonstrate that appropriately storage, security and pseudonymization techniques have been considered. In addition, it is unclear whether providing information to these users should be considered an impossible or very difficult task. \hl{In any case, the disproportionate effort it would require to provide information to the data subjects shall be demonstrated by the data controller, and is not something that should just be taken for granted.}

The concept of transparency is particularly relevant in the context of social network research, as previously highlighted e.g.~by Borgatti and Molina \cite{Borgatti2005}, and as such it requires a more extensive discussion. In particular, some additional details should provide a better description of the obligations of the data controller with regard to the provision of information. There are three points that are important here. 

First, the data controller must always provide information at the time the data is obtained (if obtained directly from the data subject) or within a reasonable period after the data is obtained and no later than a month (if the data is obtained indirectly) as long as this is possible given the appropriate adherence to the research exemptions detailed above.

Second, if the controller changes the purpose of the processing, she must provide the information to the data subject prior to this processing\footnote{Rec 61. See also Opinion where it is stated that in case the change is related to an incompatible further processing informing about the change does not "whitewash" other obligations of the controller, such as finding another lawful basis for the changed processing or asking for new consent}. For example, research data may have been collected for one purpose but the research question has shifted in the course of the data analysis and these data will now be used for a different purpose. This then speaks to how precisely the information about processing must be specified. Looking at rec 33, even though referring to consent, we can conclude that the specification in case of research can be a bit more general (such as the general research area or part of the project, not specific analytical task). Therefore changing data analysis approaches and even research questions may not require informing the data subject anew. 

Related to the above is the fact that if the change leads to further processing that is incompatible to the initial purposes, mere information of the change does not "whitewash" other obligations of the controller. According to art 5.1(b) GDPR processing should comply to the purpose limitation principle. That means that as soon as the new processing is incompatible to the initial, the controller should either avoid the new processing or find a new lawful basis for it. There is, however, an exception with regards to research purposes, since in such case the further processing for such a purpose is considered to be compatible to the initial purpose.  

Third, the general principle does not assume that the methods and the analysis are known in details at the moment of the data collection. However, the common practice in many areas of research where data is often collected with no specific hypothesis/evaluation framework becomes problematic because at least a limited explanation for the purposes of data processing is always necessary. The GDPR recognizes that it is not always possible to know from the beginning the entire scope of the research until the data is collected and used. Rec 33 (in case of consent) states that data subjects should be able to "consent only to certain areas of research or parts of research projects to the extent allowed by the intended purpose". Thus some specification of the intended purpose is necessary, limiting but not entirely eradicating exploratory forms of data collection.

\subsubsection{The depth of online social network data and the principle of data minimization}

Where some network data can be collected directly in the form of network information, that is, nodes and edges, many network datasets are obtained through processing of other types of data. For example this is often the case in research based on social media such as Twitter. Network studies of Twitter can be based on the user-articulated following/followers structure, that can be considered direct network information. \hl{At the same time, we can build networks mapping communication processes, either explicit (replies, mentions) or implicitly specified for example by the usage of common hashtags \cite{DBLP:conf/asunam/Hanteer0DM18}.} To build this second type of network, researchers collect the content of users' posts and then extract and infer \hl{relational} information. The problem arises if we consider the implications of collecting the content of the posts to build the network. Depending on the topic of posts, the type of content that is likely collected may vary but could include data revealing information that is not only identifying of natural persons but also includes sensitive data such as political affiliation, religious belief, etc. 

The GDPR makes a distinction between different types of personal data, such as data with regards to ethnicity and sexual preferences (the so-called sensitive personal data\footnote{In the context of GDPR sensitive personal data is define\hl{d} as "Personal data which are, by their nature, particularly sensitive in relation to fundamental rights and freedoms merit specific protection as the context of their processing could create significant risks to the fundamental rights and freedoms."}), and in order for the processing to be considered lawful the controller must respect the essence of data protection rights and follow suitable safeguards\footnote{Art 9 GDPR}. Notice that data which in combination with other data can lead to revealing sensitive data may also be considered as sensitive data. For example name in combination with phone number, where each piece of data is not sensitive, may constitute sensitive data together if they probably reveal the ethnicity of a person. It is easy to see how the average stream of messages written by an average user might easily contain sensitive personal data or data that can be combined to reveal sensitive personal data about the data subject. Further, such data can be derived about persons simply from information produced by their connections. For example, it may be possible to ascertain a person's political affiliation if the majority of his connections explicitly communicate theirs. 

Handling sensitive data is not forbidden, but before starting the data collection researchers need to plan some safeguards. Under the GDPR, controllers may not process sensitive personal data except if the subject has provided her "explicit consent"\footnote{Art 9(2)(a) GDPR} or the data "was manifestly made public by the data subject"\footnote{rt 9(2)(e) GDPR}, or in case of research purposes\footnote{Art 9(2)(j) GDPR}. While one may consider using the concept of "manifestly made public" for special cases such as online social networks, where the information is publicly posted online by the users, we advise against this interpretation. In fact, in the context of social media, as a consolidated body of literature has made clear, assuming when something is "manifestly public" is problematic \cite{Boyd2010} and a potentially serious breach of standard ethical research practices. On the contrary, the exemption in case of research purposes can be used, but only if processing is necessary, in accordance to Article 89(1), based on Union or Member State law which shall be "proportionate to the aim pursued, respect the essence of the data protection and provide for suitable and specific measures to safeguard the fundamental rights and the interests of the data subject." Moreover, it seems that profiling on the basis of personal data is forbidden unless there are "suitable safeguards"\footnote{Rec 51 GDPR}. For example, in Sweden, it was recommended that one such security measure can be considered the decisions of the relevant ethics committee\footnote{SOU 2017:50}. 

Finally, even if the data are not sensitive, the data minimization principle should still apply. Using again Twitter data as an example, when researchers collect information based on a hashtag they can fetch data using the hashtag with another meaning, and so not related to \hl{the} study, or data using the hashtag as was intended, but still including additional unwanted information. This means that researchers must put in place mechanisms that will effectively strip out unwanted data and delete it as soon as possible, acknowledging and documenting the process. 

\subsection{Data analysis and profiling}

Social network analysis includes a wide range of data analysis tasks. Sometimes whole-network statistics are important, for example to correlate the communication/interaction structure of a team or organization to its performance. Sometimes meso-level structures are of interest, for example if we want to identify communities \cite{Fortunato2010,Coscia2011,Bothorel2015} or other relevant sub-structures such as online conversations \cite{Magnani2012a,DBLP:journals/ans/VegaM18} inside a larger network. The identified groups can then also be used to classify individual actors, for example assigning them to a given community or role. Other types of micro-level analysis involve the characterization of single actors, for example when the most central or prestigious actors are identified \cite{Wasserman1994}. When individuals are the object of the analysis, which is the case for most of the tasks listed above, an important concept to be considered is profiling.

The GDPR puts a special emphasis on the concept of profiling by specifying the definition and codifying acceptable practices. Accordingly, in the GDPR profiling is composed of three main stages "a) collection of personal data; b) automated analysis to identify correlations; c) applying the correlation [the result of b)] to an individual to identify characteristics of present or future behaviour"\footnote{Art 29 Data Protection Working Party, WP251rev.01, "Guidelines on Automated individual decision-making and Profiling for the purposes of Regulation 2016/679"}.

Note that the notion of "automated analysis" is used in the GDPR in opposition to "manual". Although both types of processing are under the purview of the GDPR, profiling is necessarily automated. However, automated here would mean both the use of a statistical software for conducting any form of data analysis as well as the use of more complex approaches such as machine learning algorithms. Thus any data analysis that includes computational assistance from software falls under automated analysis and thus can be classified as forms of profiling. 

Given the above, many (but not all) social network analysis tasks can be classified as profiling. All centrality measures are clear examples, as they associate results of the network analysis to specific individuals. Any analysis that singles out individuals based on the identification of positions, roles and communities is similarly a form of profiling.

What is the researcher to do if their activities constitute profiling of the data subject? This does not mean that the particular data analysis is disallowed. \hl{However, this may require the performance of a data protection impact assessment (DPIA), for which the advice of the appointed data protection officer should be sought.} Although the GDPR states that profiling has to be systematic and extensive to require a DPIA, many authorities have made a broader implementation and if profiling may affect individuals in general (e.g. it provides custom access to services, it includes sensitive data, is related to vulnerable individuals, and in general the processing can lead to a high risk to the rights and freedoms of the data subject) and if it is conducted in a large scale combining sensitive data, then a DPIA is in general necessary. The question of whether a DPIA is necessary is clearly a very important one, because a very strict approach leading to an assessment for every possible case of social network analysis can become practically problematic for the researchers. While we wait for more guidelines\footnote{https://www.ucl.ac.uk/legal-services/research/data-protection-impact-assessment} and other legal specifications, the role of the researchers together with the DPOs deciding on whether an assessment is needed or not (following the law but also being practical) is of even higher importance. 

Alongside profiling, DPIAs are also applicable to systematic monitoring of individuals and locations. An interesting question arises with respect to what constitutes locations and public spaces. For example, the GDPR mentions a "systematic monitoring of a publicly accessible area on a large scale" as a reason for a DPIA\footnote{Art 35(3)(c) GDPR}. We are not aware of existing legal interpretations of whether e.g. Twitter is a publicly accessible area, but the WP29 interprets "publicly accessible area" as being any place open to any member of the public, for example a piazza, a shopping centre, a street or a public library. Clearly these are examples of physical places but Twitter is also a place that is open to any member of the public provided they have the means to access it (an internet connection and access to an email address). Such questions will likely be decided later on as the regulation stands the test of time and litigation, but it is an important item to consider for researchers conducting large-scale collection and processing of ostensibly "public" data.

\subsection{Data storage}

In this section we discuss what happens after the research is concluded, in case the researchers want to store the collected networks. If the data are still personal, e.g., they still contain identifiers or have been pseudonymized, then the data controller must guarantee some rights to the data subjects if she wants to keep the network data. On a general level we can organize these rights along three lines: a) temporal duration of personal data storage, b) the accessibility of the stored data to the data subject, c) the right of the data subject to withdraw his/her data. All these tasks are in general strictly regulated by the GDPR, but with significant exemptions for research, discussed in the following section. Under the assumption that the networks have been anonymized, then there is no problem because the GDPR no longer applies: the data are no longer personal. However, network anonymization is a complex issue, that we also discuss below.

\subsubsection{Rights of the data subjects and the principle of storage limitation}

When it comes to temporal storage limitation, the GDPR states that in general data can be "kept in a form which permits identification of data subjects for no longer than is necessary for the purposes for which the personal data are processed", but more extended periods may apply in case of research as long as appropriate safeguards are followed\footnote{Art 5(e) GDPR}.

No exemption because of research is instead mentioned regarding the data subjects' right to check if there is data concerning them, and the right to obtain these data\footnote{Art 15 GDPR}. This means that when requested the controller should provide the data, in a "commonly used and machine-readable format"\footnote{Art 20 GDPR} (even if there are possibly other national laws that may restrict this right of a data subject, such as for example secrecy acts\footnote{SOU p. 223}). Considering the average amount of data represented by a single node in a typical social network project, this should not be a problem. Nevertheless, as for other parts of this article, the size of the network may constitute a practical difference, and for large networks researchers should probably consider implementing an automated data filtering functionality.

Finally, the right to erasure, also known as right to be forgotten, grants to the data subject "the right to obtain from the controller the erasure of personal data concerning him or her without undue delay"\footnote{Art 17 GDPR}.  However, also in this case the GDPR contains an exemption to this obligation if the erasure "is likely to render impossible or seriously impair the achievements of the objectives of that processing"\footnote{Art 17 GDPR}. Many SNA measures are not so sensitive to a small amount of missing data \cite{kossinets2006effects} and the discipline has developed a set of techniques to handle missing data. Nevertheless, it should be acknowledged that a significant number of subjects requesting their data to be removed might seriously impair the research objectives, thus researchers would have the right to legally object to the data removal.

While these are the general guidelines emerging from the GDPR, according to art 89(2) Member States may further limit the data subjects right to access, rectification, restriction and to object in case of research if there are appropriate safeguards in place, and as long as the derogation is necessary for the fulfillment of the research.

\subsubsection{Data anonymization}

The GDPR asks for appropriate safeguards. The safeguards that are named in the GDPR are technical and organizational, e.g. data minimization, pseudonymization and anonymization. In addition, there can also be legal safeguards, such as contractual clauses between the controller and the processor, ethical vetting etc. \cite{MagnussonSjoberg2017}. Here we focus on anonymization, which should result in the data not being re-identifiable by the controller or any other person. In social network analysis, the typical approaches to anonymization are based on clustering, graph modification or network perturbation \cite{Zhou2008}. 

Data anonymization approaches in general are part of a considerable debate where some researchers argue that anonymization is impossible while others contend that it is in some cases \cite{Cavoukian2011,Narayanan2014,Narayanan2016}. Social network data is far more difficult to anonymize than other types of data and research on appropriate anonymization techniques is still in its relative infancy. Many of the simpler and more traditional approaches such as replacing node identifiers as well as more recent and complex approaches have been critiqued as insufficient \cite{Backstrom2007,Narayanan2014}. The knowledge of research being conducted in a particular location by a specific research group may be enough to reveal the identities of individuals encoded in the network to those who are familiar with these people more directly. In a small social network, such as for example a company division, it may be simple for the people in the network to recognize others based on just the revealed relational patterns \cite{Borgatti2005}. Such an issue is not specific of social networks, but has been amply documented in qualitative and ethnographic research \cite{Shklovski2013,VandenHoonaard2003}. As another case, if the data are public and indexed (e.g., by Web search engines), it can be very easy to find the original data using a part of it as a search key, such as finding the authors of a social media post based on the text of the post. 

Whether anonymization or even just pseudonymisation are generally possible in a social network context is a difficult question. The GDPR states the necessity for privacy by design and by default but does not request specific privacy-preserving solutions: the controller should select and apply the appropriate measures for each case. In the GDPR, pseudonymisation requires the "additional information" to be "kept separately" and to be "subject to technical and organisational measures"\footnote{Art 4(5) GDPR}, which is not really possible when the data source is public: if one removes the user identifier but keeps the text of the post (e.g., the tweet), a simple search on a search engine or on the social media platform can easily lead to the original, complete information. In this case, a possibility to be considered by the researchers (but not explicitly required by the GDPR) is to transform the text so that the analysis can still be performed but it becomes more complicated to fetch it from the Web, such as replacing it with a bag of words. The relevance of this discussion is that according to rec 26 pseudonymized data is identifiable, so the GDPR applies to that, while anonymized data is not, so the GDPR is no longer relevant. However, given the difficulty in fully anonymizing the data we should often assume that the GDPR is still the relevant regulation.

Even when we do not need identifiers to process social network data, because for example we are only interested in the structure of the network and its relationship with some indicators, we still need the identifiers if we want to extend the network, to know to which nodes the newly available information refers to. According to the GDPR we should at the very least pseudonymise the data "as soon as possible" (recital 78). However, it is not unusual in online network studies to keep collecting data for months or even years, which means that "as soon as possible" may be as late as the end of the study. One solution here is to develop or extend data collection systems with built-in network pseudonymisation functions, for example automatically removing identifiers and separately storing a mapping to user accounts in a location that requires special access credentials. Such solutions may seem overly onerous given the current accepted practices, but the GDPR forces us to rethink our attitudes towards data collection and the impacts of our practices more broadly. In addition, the idea of designing ethically-related features in social network analysis software has already appeared in the literature \cite{Borgatti2005}.

As a final note, while in the previous paragraphs we have discussed the difficulty of network data anonymization, there are specific types of social network data where anonymization is indeed possible. In ego-network data collection different actors are asked about their own social ties and perhaps those of their neighbors. Ego-networks are then analysed without reconstructing a common network for all of the participants. In this case, there is typically no need to know who the individuals are, which means that we can design a data collection that is already anonymised at the source. As a result, these data are outside the scope of the GDPR given the definition of anonymous data as "data rendered anonymous in such a way that the data subject is not or no longer identifiable"\footnote{Rec 26 GDPR}.

\section{\hl{Flows of data}}

\hl{So far we have considered cases where only one data controller processes the data, as in the case of a single research team based at one institution performing the research. In practical situations it can however happen that data is stored by a team at a university and sent to a team at another university to be analyzed, or that two universities perform a joint data collection.}

\hl{It is not the goal of this paper to examine the legal implications regarding data flows from one jurisdiction to another but there are some things that are worth naming here. Firstly, when it comes to transfers of personal data within the EU, the GDPR has as a goal to "prevent divergences hampering the free movement of personal data within the internal market"\footnote{Rec 13 GDPR}. However, when it comes to data related to research the situation is somewhat more complicated since quite many issues are left to the Member States to decide\footnote{Art 89 GDPR. For a short analysis on the matter see also Staunton C, Slokenberga S and Mascalzoni D, 'The GDPR and the Research Exemption: Considerations on the Necessary Safeguards for Research Biobanks' [2019] European Journal of Human Genetics, http://www.nature.com/articles/s41431-019-0386-5 accessed 18 June 2019.}.}

\hl{Moreover, regarding transfers to third, non-EU, countries, it has to be made sure that such transfers comply to the safeguards provided for in art 44 et seq  GDPR. We will not analyze the different possibilities for such compliance to be achieved but it is important to emphasize that when it comes to transfers of data to entities in third countries the situation is far from problem-free. By way of illustration such transfers are allowed if the data are sent to a "safe country", namely to a country recognized as a country providing an equally adequate level of data protection as the EU countries. However for the time being these countries are limited to a handful of -- mostly minor -- countries\footnote{Andorra, Argentina, Canada (commercial organisations), Faeroe Islands, Guernsey, Israel, Isle of Man, Jersey, New Zealand, Switzerland, Uruguay and the USA but only with respect to the Privacy Shield-certified
companies. See https://ec.europa.eu/info/law/law-topic/data-protection/international-dimension-data-protection/adequacy-decisions\_en.}. Alternatively, such transfers must be based on the consent of the data subject, something which, however, as already stated above can be difficult to be obtained in cases of research in networks. Similarly, transfers to third countries are allowed if the data-transferring party and the data-receiving party  use an EU Standard Contractual Clause. However such clauses have already been challenged with regards to their ability to provide an adequate level of protection of personal data\footnote{See Reference for a preliminary ruling from the High Court (Ireland) made on 9 May 2018 -- Data Protection Commissioner v Facebook Ireland Limited, Maximillian Schrems
(Case C-311/18).}.}

\section{Some more general issues and considerations}

\subsection{A code of conduct for social network research}

The opportunity of writing a code of conduct for research in social network analysis has been under discussion for a long time. In the special issue of Social Networks on ethical dilemmas in social network research, there was a mention to "efforts now underway within INSNA, the professional association for social network researchers, to establish a code of ethics" \cite{Breiger2005}. Several years later, at a board meeting of the same association\footnote{2016 Report to INSNA Membership prepared by the INSNA Officer.} this was still under discussion, and it was noted that many members of the association are also members of other associations for which codes of conduct already exist (e.g., by the American Anthropologist Association, the American Political Science Association, the American Sociological Association, the \hl{Association for Internet Researchers -- AOIR}), questioning the need for an additional effort.

We believe that this article can contribute to this discussion. On the one hand, we note that many issues highlighted in the previous sections are common to other types of research not necessarily involving social networks, including for example social science research in general, Internet research and big data analysis, even though some specific aspects of social networks have also emerged and the combination of relevant issues is also unique. On the other end, the broad picture emerging from our analysis of the GDPR is a complex one, and a whole section of the regulation\footnote{Sec 5, art 40-43} indicates codes of conducts as a way to reduce this complexity. In fact, once a code of conduct proposed by an association has been approved, registered and published by a supervisory authority certifying that the code is compliant with the GDPR and "that it provides sufficient appropriate safeguards"\footnote{Art 40(5)}, then showing compliance with the code exempts the data controller from a number of obligations. In summary, after the enforcement of the GDPR the benefits of codes of conducts have increased, but their establishment requires additional effort because they require an authority to verify their compliance. 

\subsection{Towards GDPR-compliant social network software}

Through the analysis of the legal obligations emerging from the GDPR we have seen many cases where the law can be considered a bottom line for ethics, where individual researchers shall consider more restrictive actions. For example, as we have discussed above, the GDPR explicitly mentions "disproportionate effort" as a reason not to provide information to the data subjects. This, when framed within the context of online data or of secondary analysis of already collected large datasets, might easily be used as a solid reason to perform research without informing the data subjects. But if it is true that large online data could easily count millions of potential data subjects, one can expect that for online sources it can be possible to automatically send notifications or messages informing the data subjects. While this may result in a potentially significant overhead of communicating with confused data subjects, the effort may be a first step in acknowledging that people that produce data must be treated with dignity and respect regardless of research aims. Development of standards for notification in large-scale data collection endeavours is necessary and may need to be taken up at the level of professional codes of conduct.

In these cases, we should also consider developing tools that can take care of notification automatically, reducing the claim of disproportionate effort rather than leveraging it as a way to side-step responsibilities in research. For example, in the growing context of Twitter research, sending a short tweet mentioning those user accounts included in the social network data collected in a research project would be potentially interesting information for data subjects, contributing to the creation of an awareness about how much our public data is used. If done by a relevant share of researchers (which can theoretically be achieved if the main tool or tools for data collection are extended with this functionality) this increased awareness could result in a consequential generalized improvement in the way people manage their data online and an increased trust in science, showing how careful researchers are about this. However, while automatically sending the information to a list of users seems to require a limited effort, turning this into practice can be problematic, as described in the next section.

\subsection{An experiment on automated data subject information on Twitter}

To better understand the amount of effort needed to notify data subjects in the context of online social network research we have set up a protocol for a Twitter data collection process. This experiment, briefly reported below, highlighted the difficulty of performing even a task (appearing to be) as easy as sending some information to online users.

First, we had to consider a number of alternatives. First, when tweets are collected on Twitter the only contact information we have are the Twitter identifier and screen name of the accounts whose tweets were collected. This means that we can only inform Twitter users via Twitter, and as it is typically not possible to send direct (private) messages to generic Twitter users (for example, users not following us who have not explicitly allowed this in their privacy settings) we need to inform them in some way that is visible to others, such as using a public mention.

While this is not necessarily problematic, it is interesting to see how to inform data subjects so that we can protect their privacy we have to release additional information about them: our public message implies that those accounts have posted tweets with the hashtag we were monitoring, that by the principle of transparency we have to clearly indicate in the communication. Then, we must decide (1) how many users to mention in the same tweet and (2) whether we should check what their current screen name is. Both choices have an impact on the time needed to send the notifications: including more accounts in the same Tweet would reduce notification time, but would also again release more information as each notified user would see the other user names in the same tweet, knowing that they have also used the same hashtag. Checking the current screen name would require more accesses to the Twitter API but would avoid that we mention the wrong account because screen names can change in time. As an indication, the Twitter API currently allows us to send 2400 tweets per day, meaning that we would need around one year and two months to notify one million users (using a single notification account).

An alternative is to notify users through the hashtag, sending a tweet without mentions but with the monitored hashtag and specifying that we are collecting tweets containing the same hashtag. This is however also problematic, first because there is no guarantee that users will see that (they would have to search tweets containing that hashtag, at the right time), second because for some studies awareness of the data collection might result in a different behaviour.

Other decisions making the practical information process less trivial than one may think are whether we should also notify accounts mentioned in the collected tweets, even if they were not producing tweets themselves, or whether we should notify accounts retweeting other accounts' tweets.

After deciding on all these aspects, we started sending our tweets, including a link to the information about the project and the data processing and also information about the user rights. The procedure for the users to offer them (among other things) the possibility of retrieving the data about them we had collected was also complicated, because to prove their identity the users were requested to follow our notification account - which is again revealing more information about the user and also requires some effort that might discourage potentially interested users. After sending notifications to 45 accounts we registered only one visit to the information page.

Finally, Twitter blocked our notification account. According to their rules, the account had been marked as having a spamming behaviour. In the process to reactivate the account we mentioned that despite the behaviour being compatible with their definition of spamming, the account was an attempt to enforce the rights of the users to know that their tweets had been collected and why, but this did not result in any exception, which led us to drop the experiment after considering that developing a "smarter" bot sending the notifications and trying to behave in a way not to be caught by Twitter's algorithms would have been ethically questionable.

\subsection{Public vs.~private actors}
\label{sec:private/public}

Another important point of discussion is the difference in classification of universities and commercial research and industry labs. The stark difference in legal basis for data processing and the impact on the consideration of whether consent is a legitimate lawful basis is an important point to consider. What does consent constitute in the context of a commercial entity when it must be clearly uncoerced and freely given? What are the different obligations towards data subjects for researchers depending on the legal ground they employ for data collection? These questions have some answers in the GDPR but will be further evolving as time and litigation test the GDPR terms and definitions. Data availability for large scale computational social science and social network research is necessarily connected to commercial actors \cite{Lazer2009}. Collaborators across academic and commercial spheres have claimed the unalloyed public good that is possible from large-scale data collection, but what impact may the GDPR have, given the differentiation it makes between public and commercial research efforts? How much access will public university researchers continue to have to commercial data stores? How complex will these negotiations become? These questions are beyond the purview of this paper, but must be discussed and considered in the future.

\section{Conclusions}

Our main objective when we started writing this article was to provide a practical guide to GDPR-compliant social network data processing. Working on it, and also trying to apply our recommendations to our own research, it became evident that while some issues could be more easily translated into practical suggestions, other general indications and principles in the regulation are difficult to either interpret or apply in the context of social network research. The problems we have highlighted in the article include the difficulty of sending information to millions of users through a third-party API that does not allow it, the problems in pseudonymizing the data as soon as possible in a continuous network monitoring process performed with pre-GDPR software tools, the interpretation of concepts such as "manifestly made public data" and "publicly accessible areas", the problem of removing data by user request not knowing what impact this will have on network statistics, the practical impossibility of guaranteeing respondent anonymity, the inclusion of data about individuals not included in the study, as well as more general issues related to data protection emerging when social network analysis is applied to large-scale networks of social relations derived from social media data.

In summary, it is important that everyone involved in the processing of social network data invests some time to reflect about the implications of the GDPR on their research, seeking help from their institutions but not only relying on institutional support. While this may sound as an obvious statement, and legal and ethical problems related to social network analysis and Internet research for social behaviour have certainly received a lot of attention in the past as witnessed by the literature on the topic and by existing codes of conduct, the sudden explosion of online behavioural data has indeed affected the research landscape by both introducing new problems and involving new researchers from disciplines \hl{where} some of these problems had not been traditionally accounted for.

\section{Acknowledgments}

This work was partially supported by the European Union's Horizon 2020 research and innovation programme under grant agreements No. 727040 (Virt-EU).

\bibliographystyle{plain}
\bibliography{sna,ml}

\end{document}